\documentclass[cameraready]{Interspeech}
\usepackage{comment}
\usepackage{cite}
\usepackage{tipa}
\usepackage{verbatim}
\usepackage{makecell}
\usepackage{amsmath,amssymb,amsfonts}
\usepackage{algorithmic}
\usepackage{graphicx}
\usepackage{textcomp}
\usepackage{xcolor}
\usepackage{tipa}
\title{Speaker Verification Under Real Classroom Conditions for English Speech}

\author[affiliation={1}]{Saba}{Tabatabaee}
\author[affiliation={2}]{Jing}{Liu}
\author[affiliation={2}]{Meghavarshini}{Krishnaswamy}
\author[affiliation={1}]{Carol}{Espy-Wilson}

\address{
    $^1$ Department of Electrical and Computer Engineering, University of Maryland College Park\\
    $^2$ Center for Educational Data Science and Innovation, University of Maryland College Park\
}

\email{sabatb@umd.edu, jliu28@umd.edu, mkswamy@umd.edu, espy@umd.edu}


\usepackage{comment}

\begin{document}

\maketitle

\begin{abstract}
Developing speaker verification (SV) models that are robust to classroom noise and effective across both children and adult speakers is critical for AI tools supporting educational environments. In this study, we use a real-world English-speaking classrooms dataset containing partial speaker identity annotations, with most recordings remaining unlabeled. We adapt the WavLM-TDNN model for classroom SV, achieving average relative reductions in Equal Error Rate (EER) of 23.99\% and 6.32\% compared to the ECAPA-TDNN baseline and the ECAPA-TDNN model trained on classroom data, respectively. 
Additionally, we investigate two training strategies for SV in classroom settings: self-supervised learning (SSL)  and a two-stage approach that first pre-trains with SSL and then fine-tunes with limited annotated data. Five-fold cross-validation demonstrates that the two-stage strategy consistently outperforms the SSL-only approach, achieving an average relative EER reduction of 13.39\%.
\end{abstract}

\section{Introduction}
Robust speaker verification (SV) systems for classroom environments are essential for enabling intelligent educational technologies that facilitate the monitoring of students participation and teacher–student interactions. However, most existing SV research has focused primarily on adult speech \cite{kim2024self, li2025disentangling,gan2025idir,han2025noise}, with comparatively limited attention to classroom environments that include both children and adults \cite{tabatabaee25_interspeech,Zheng2024NResNetNR,Kadyan2022AutomaticSV}. 

Models trained on adult datasets often generalize poorly to children’s speech due to acoustic characteristics differences. This challenge is further amplified in classrooms, where both children and adults are present under unique acoustic conditions and background noise, such as babble, making SV particularly difficult. Another major challenge in developing SV systems for classroom settings is the limited availability of publicly accessible speech datasets recorded in real classroom environments. To address this gap, we leverage real-world audio collected from 6th to 8th grade English-speaking classrooms to develop and evaluate SV tasks under realistic classroom conditions.

Previous SV methods have relied on supervised learning (SL), which maps an input utterance to its corresponding speaker identity using labeled training data. While effective, these approaches are limited by the scarcity of annotated classroom recordings and the significant time and cost required to obtain high-quality labels, restricting their practical deployment in real-world settings. To address these limitations, self-supervised learning (SSL) has emerged as a promising alternative. SSL frameworks learn informative and discriminative representations directly from raw audio without the need for annotations. Among SSL approaches, contrastive learning methods such as momentum contrast (MoCo) \cite{he2020momentum} and non-contrastive methods like DINO \cite{caron2021emerging} are commonly used for SV tasks \cite{han2023self, tu2024contrastive}. A study \cite {garrido2022duality} has shown that contrastive and non-contrastive learning objectives are closely related, with their optimization is being equivalent up to row and column normalization of the embedding matrix. This finding suggests that, given careful tuning of model architectures, loss functions, and hyperparameters, contrastive and non-contrastive methods can achieve comparable performance in SV tasks. 

Given that the majority of our classrooms data is unannotated, we adopt MoCo to learn speaker embeddings directly from these recordings in a self-supervised manner. Furthermore, we employ a two-stage training approach, first pre-training the embedding model using the self-supervised MoCo framework, and then fine-tuning it on a limited set of annotated classrooms data. This two-stage training approach is evaluated and compared with the SSL-only approach to assess the impact of supervised adaptation on SV performance in classroom settings.

%In the MoCo SSL framework, utterances from different audio files are assumed to belong to different speakers, and the model is trained contrastively to maximize similarity between positive pairs while minimizing it between negative pairs. 

With the rise of deep learning, traditional SV approaches, such as i-vectors \cite{dehak2010front}, have been increasingly surpassed by deep neural network-based methods, including x-vector architectures \cite{snyder2018x} and, more recently, the ECAPA-TDNN framework \cite{desplanques2020ecapa}. Recent self‑supervised pre‑trained models such as wav2vec 2.0\cite{baevski2020wav2vec}, HuBERT\cite{hsu2021hubert}, and WavLM \cite{chen2022wavlm} have reshaped speech processing by replacing conventional features like filterbanks with learned representations that capture richer speaker‑discriminative information, leading to superior performance in downstream SV tasks\cite{chen2022large, sankala2022multi, yang2021superb}.

Previous works have explored combining pre-trained speech models with time delay neural network (TDNN) backends for SV \cite{novoselov2022robust,kim2024layer}. Novoselov et al. \cite{novoselov2022robust} proposed a wav2vec 2.0–TDNN model, demonstrating improved robustness and cross-domain generalization compared to the ECAPA-TDNN framework. Moreover, Kim et al. \cite{kim2024layer} used WavLM as a frontend with TDNN as the backend and reported superior SV performance over other pre-trained models, including HuBERT and wav2vec 2.0, when combined with TDNN. However, these studies have primarily focused on adult speech and have not evaluated performance in environments containing both children and adults, such as classrooms, which present unique acoustic challenges. To address this gap, we propose a WavLM–TDNN model that leverages WavLM for audio representation extraction and TDNN layers as the backend, and we evaluate its performance against ECAPA-TDNN in classroom SV tasks. To the best of our knowledge, this is the first study to investigate the use of a pre-trained speech model for SV in classroom settings.

\textbf{Our key contributions} in this work are as follows:
\begin{itemize}
    \item Using a real-world dataset of 6th to 8th grade classrooms, including both enrollment and classrooms recordings.
    \item Comparing SSL-only training with a two-stage training approach for SV in classroom environments.
    \item Comparing the performance of the proposed WavLM-TDNN model with the ECAPA-TDNN model for SV in classrooms.
    \item Extending the evaluation of the proposed WavLM-TDNN model on the SV task by using a multi-lingual classrooms dataset and comparing its performance with the ECAPA-TDNN model.
\end{itemize}
 %\vspace{-1mm} 
\section{Methodology}
\subsection{Dataset description}
%developed by the Center for Education Data Science and Innovation (EDSI)
We used an in-house multi-modal classrooms dataset, referred to as the EDSI dataset. EDSI consists of audio and video recordings from 6th to 8th grade mathematics classrooms, along with relevant student information such as demographic information, student achievement data, seating charts, and psychometric data collected through student and teacher surveys. Each classroom contains a distinct group of students, and there is no speaker overlap between classrooms. 

Classroom audio was captured using five Swivl M2 microphones \footnote{https://swivl.zendesk.com/hc/en-us/articles/27423075821979--M2-Specifications-Microphone-and-Speaker}. One microphone was worn by the teacher and four microphones were placed on desks across the room to capture student speech. Audio signals from multiple microphones were averaged to produce a single unified signal for analysis. We randomly selected 18 English-speaking classrooms, comprising a total of 316 sessions, each session lasting approximately one hour. For each classroom, a voice enrollment session was conducted, during which both teachers and students individually read a short passage lasting approximately 30–60 seconds. Across the 18 classrooms, the enrollment data comprise a total of 4.31 hours of speech from 402 speakers, with one recording per speaker. The enrollment data were used exclusively for evaluating SV model performance and were not included in model training.

Overall, the 18 classrooms dataset comprises around 218 hours of speech data and 108714 utterances. To generate ground-truth speaker IDs, one session from each classroom was randomly selected. Two trained annotators assigned speaker labels to every utterance using the unified audio signal, along with contextual information from the seating chart, class roster, and classroom video. The speaker annotation task resulted in ground-truth information for 10.22 hours of speech, 4017 utterances, and 216 unique speaker IDs. This portion of the dataset is referred to as EDSI (W-ID), where W-ID denotes with speaker ID. The remaining sessions, which do not have speaker IDs, comprise 207.55 hours of speech and 104697 utterances, and are referred to as EDSI (WO-ID), where WO-ID denotes without speaker ID. To ensure speaker independence during evaluation, data splitting was performed at the classroom level. Five-fold cross-validation was conducted by randomly selecting four classrooms from EDSI (W-ID) as the test set in each fold, with data from the remaining 14 classrooms, including both EDSI (W-ID) and EDSI (WO-ID), used for training.

 \vspace{-1mm} 
\subsection{Developing proposed WavLM-TDNN model}
As shown in Figure 1, the WavLM-TDNN model combines TDNN with WavLM-Large, leveraging WavLM’s contextualized speech representations and TDNN’s ability to capture both local and long-range temporal patterns. The input speech signal is first processed by WavLM-Large to extract frame-level representations from each of its 25 transformer layers. These 25 layer-wise representations are then combined using a learnable weighted sum normalized with softmax to produce a single aggregated representation. This representation is subsequently passed through two separate dense layers, each containing 512 units. 

The output of the first dense layer is passed through a stack of five TDNN blocks with dimensions (512, 512, 512, 512, 1500), kernel sizes (5, 3, 3, 1, 1), and dilation rates (1, 2, 3, 1, 1). Each TDNN block performs a linear transformation over a temporal window followed by a ReLU activation. By employing dilated convolutions, the TDNN captures both short-term and long-term temporal patterns in speech, effectively modeling speaker characteristics across multiple time scales. The final TDNN output is projected into a lower-dimensional space via a dense layer with 256 units and then aggregated using attention pooling. Attention pooling computes frame-wise scores, normalizes them with a softmax across time, and produces a weighted sum to generate an utterance-level representation.

The output of the second dense layer with 512 units is passed through a gaussian error linear unit (GELU) activation, followed by a dense layer with 256 units, and subsequently processed with temporal attention pooling, yielding a second utterance-level representation. This representation is fused with the TDNN-derived representation using a learnable weighted sum to form the final 256-dimensional speaker embedding, effectively combining TDNN-based temporal modeling with self-supervised WavLM-Large features. During training, the WavLM encoder is not frozen, enabling joint optimization of the entire architecture and allowing both the WavLM backbone and TDNN layers to adapt to the SV task.

\begin{figure*}[htbp]
    \hfill
    %\vspace{-3mm}
    \includegraphics[width=1\textwidth, height=0.24\textheight]{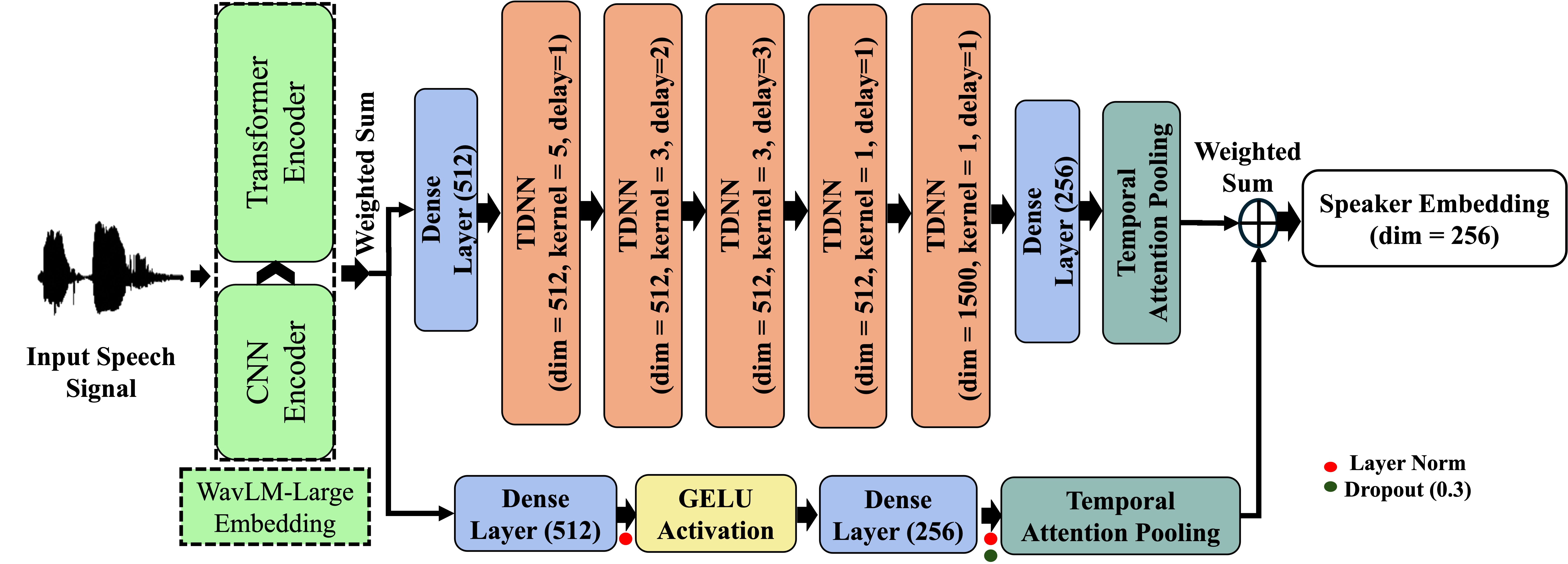}  % Adjust width as needed
    \caption{Proposed WavLM-TDNN model architecture.}    
    \label{fig:si}
    %\vspace{-2mm} 
\end{figure*}
 \vspace{-1mm} 
\subsection {Training methods}
\subsubsection{Self-supervised learning method}
We adopt a self-supervised contrastive learning framework based on MoCo to learn speaker-discriminative representations from the EDSI corpus. The objective is to train an embedding function that maximizes agreement between augmented views of the same utterance while minimizing similarity across different utterances. The MoCo framework uses two encoders with identical architectures: a query encoder $f_q$ and a key encoder $f_k$. The query encoder is optimized via backpropagation, while the key encoder is updated as an exponential moving average (EMA) of the query encoder parameters with momentum $m = 0.996$. Denoting the parameters of $f_k$ as $\theta_k$ and those of $f_q$ as $\theta_q$, the update rule is shown in Equation~\ref{eq:momentum_update}.
\begin{equation}
\theta_k \leftarrow m \theta_k + (1 - m) \theta_q
\label{eq:momentum_update}
\end{equation}
This update stabilizes the key encoder by producing slowly evolving target representations and prevents rapid fluctuations. Nesterov momentum is employed during the optimization process to further improve convergence. 
 The model is optimized using the InfoNCE objective as shown in Equation~\ref{eq:loss}.
 
\begin{equation}
\mathcal{L}_{\text{SSL}} =  \frac{1}{N} \sum_{i=1}^{N} 
-\log \frac{\exp \left( x_{q_i} \cdot x_{k_i}^+ / \tau \right)}
{\sum_{j=0}^{K} \exp \left( x_{q_i} \cdot x_{k_{i,j}} / \tau \right)}
\label{eq:loss}
\end{equation}

where $x_{q_i}$ denotes a query sample (i.e., anchor) and $x_{k_i}^+$ is its corresponding positive key. The negative keys $x_{k_{i,j}}^-$ are encoded by the key (momentum) encoder. In Equation~\ref{eq:loss}, the sum is over one positive pair and K negative pairs, with $\tau = 0.07$ as the temperature hyperparameter. To increase the number of negative samples beyond a single mini-batch, we employ a memory queue of size K = 65536, which stores key embeddings from previous iterations in a first-in-first-out manner, dequeuing the oldest embeddings as new ones are enqueued.

Positive pairs are generated through speech augmentations, including additive noise and reverberation with a probability of 0.6 per training sample, along with random chunk sampling. These augmentations encourage the encoder to focus on speaker-specific features rather than channel or environmental cues. For noise augmentation, we use non-babble environmental noise from the DNS Challenge dataset \cite{reddy2021interspeech} and babble noise generated from the My Science Tutor (MyST) children dataset \cite{Pradhan2023MyST} by randomly overlapping speech from 5 to 20 speakers. The training signals are augmented with randomly selected SNRs ranging from 0 to 15 dB. Reverberation is simulated using impulse responses from OpenSLR\footnote{https://www.openslr.org/28/}. Training uses stochastic gradient descent (SGD) with momentum 0.9 and weight decay 1e-4. The model is trained for 250 epochs with an initial learning rate of 5e-3, linearly warmed up from zero over the first ten epochs, followed by exponential decay to 5e-5. 
\subsubsection{Two-stage training method}
In this study, we employed a two-stage training strategy to compare its performance with an SSL-only approach (see Section 2.3.1) and assess its effectiveness for SV in classroom environments. The two-stage training approach consists of two steps: first, pre-training the model using the SSL method on the EDSI dataset, including both EDSI (WO-ID) and EDSI (W-ID); and second, fine-tuning the model with supervised learning on the EDSI (W-ID) dataset. This approach allows the model to leverage rich representations learned from unlabeled data before adapting to the supervised SV task.

In supervised training, additive noise and reverberation, along with random chunk sampling, are applied to each sample with a probability of 0.6, following the SSL augmentation procedure described in Section 2.3.1. A projection layer with an output size equal to the number of speakers is appended to the end of the embedding model to enable classification.

The supervised training loss combines multiple complementary components, as shown in Equation~\ref{eq:loss_sl}. Cross-Entropy Loss (CE) applied on the projection layer guides the model to correctly classify speaker identities, along with AAM-Softmax (ArcFace) and a triplet margin loss, which promote intra-speaker compactness and inter-speaker separation. Semi-hard negative mining is used to select more challenging negative examples to enhance the effectiveness of the triplet loss.

\begin{equation}
\mathcal{L}_{\text{SL}} = \mathcal{L}_{\text{CE}}+\mathcal{L}_{\text{AAM-Softmax}} + \mathcal{L}_{\text{Triplet}}
\label{eq:loss_sl}
\end{equation}

The model is optimized using SGD with a momentum of 0.9 and a weight decay of 1e-4, along with Nesterov acceleration to improve convergence. During SSL pre-training, the model is optimized with an initial learning rate of 5e-3, which is exponentially decayed to 5e-5. For supervised fine-tuning, the learning rate is initialized at 5e-5 and exponentially decayed to 5e-7. 

SV performance is evaluated by pairing enrollment data with the EDSI (W-ID) classroom test set and computing the Equal Error Rate (EER) from the resulting verification trials.
 %\vspace{-1mm} 
\section{Results and discussion}
%Since the enrollment data were recorded in mostly quiet conditions while the test data were collected in noisy classroom environments, the EER reflects both the model’s speaker discrimination ability and its robustness to environmental noise and acoustic variability.

\subsection{Comparison of the training methods for the WavLM-TDNN}
Table 1 compares the SSL and two-stage training strategies applied to the WavLM-TDNN model across five folds, evaluated using EER. The two-stage approach consistently outperforms the SSL-only method across all five folds, achieving an average relative reduction of 13.39\% in EER. These results demonstrate that incorporating even a small amount of labeled speaker data improves SV performance in classroom scenarios. 

Our analysis of each classroom’s data and their corresponding EERs in each fold suggests that, although classroom recordings are treated as a single domain, specific classroom settings, whether collaborative or instructional, affect factors such as background noise, which in turn affect the EER. We observe that some teachers encourage students to work in groups and engage in discussions, which often leads to higher levels of babble noise and creates a more challenging acoustic environment. Folds containing a larger proportion of these group discussion recordings, such as Fold 5, tend to be more difficult for the SV task, resulting in higher EER. In contrast, folds with fewer group discussions, such as Fold 2, exhibit lower EER due to the reduced babble noise.
\begin{comment}
\begin{table*}[htbp]
\vspace{-1mm} 
\caption{Comparison of audio embedding models for speaker verification.}
\footnotesize
\setlength{\tabcolsep}{7pt}
\renewcommand{\arraystretch}{1.3}
\centering
\begin{tabular}{c|c|c|ccccc|c}
\hline
 Method & Training data & Fine-tuning data &Fold 1 %(Hawkins & Zazycki &Snipes &Cacaci)  
 &Fold 2 %Leggett&Cacaci & Kerr&Hawkins 
 &Fold 3 %Loux&Leggett&Paul&Snipes
 &Fold 4 %Snipes&Zazycki&Legget  &Britt
 &Fold 5 %Kerr&Hawkins&Crigger  &Paul
 &Average of Folds 
 \\
\hline
\multicolumn{9}{c}{\hspace{3.8cm}  ECAPA-TDNN (Base) } \\ \hline 
- &Voxceleb &- &17.51 &18.80&19.48&21.75&21.69&19.85\\
\hline
\multicolumn{9}{c}{\hspace{3cm}  ECAPA-TDNN } \\ \hline 
SSL&EDSI (WO-ID)&- &15.09  &14.57&16.32&18.53&19.74&\\
SL&EDSI (W-ID)&-&14.55 &17.11&17.43&&&\\
Hybrid&EDSI (WO-ID) &EDSI (W-ID)&12.94 &16.81&15.08&17.86&17.31&\\\hline

%\multicolumn{6}{c}{\hspace{2cm} WavLM-MHFA} \\ \hline
%SSL&EDSI (WO-ID) &-&14.59 (2.42)&&&&&\\
%SL&EDSI (W-ID)&-&13.65 (1.02)&&&&&\\
%SSL-SL(FT)&EDSI (WO-ID)& EDSI (W-ID)&11.31 (2.10)&&&&&\\\hline
\multicolumn{9}{c}{\hspace{4.1cm} Proposed WavLM-TDNN} \\ \hline
SSL&EDSI (WO-ID)&-&14.91&17.70&18.28&&&\\
SL&EDSI (W-ID)&-&12.61 &17.00&16.79&17.16&&\\
Hybrid&EDSI (WO-ID) &EDSI (W-ID)&\textbf{11.84}&16.01&\textbf{14.14}&&&\\\hline

    \end{tabular}
    \vspace{-1mm} 
  \end{table*}
\end{comment}

%fold 1 is fold3
%fold 2 is fold 3

  \begin{table}[htbp]
%\vspace{-1mm} 
\caption{Comparison of training methods for developing the proposed WavLM-TDNN model across five folds using EER (\%). Avg: average EER across five folds. The best results are shown in bold.}
\footnotesize
\setlength{\tabcolsep}{3pt}
\renewcommand{\arraystretch}{1.4}
\centering
\begin{tabular}{|l|ccccc|c|}
\hline
 Method &Fold 1 %(Hawkins & Zazycki &Snipes &Cacaci)  
 &Fold 2 %Leggett&Cacaci & Kerr&Hawkins 
 &Fold 3 %Loux&Leggett&Paul&Snipes
 &Fold 4 %Snipes&Zazycki&Legget  &Britt
 &Fold 5 %Kerr&Hawkins&Crigger  &Paul
 &Avg
 \\
 \hline 
SSL&18.28&14.91&17.70&18.74&19.28&17.78\\\hline
%SL&16.79&12.61 &17.00&17.16&19.38&\\\hline
Two-stage&\textbf{14.14}&\textbf{11.84}&\textbf{16.01}&\textbf{16.62}&\textbf{18.41}&\textbf{15.40}\\\hline

    \end{tabular}
    %\vspace{-3mm} 
  \end{table}
  \subsection{Comparison of the WavLM-TDNN with the ECAPA-TDNN}
Table 2 compares the proposed WavLM-TDNN model with the ECAPA-TDNN (base) model and an ECAPA-TDNN model trained on the EDSI dataset using the two-stage training approach. The ECAPA-TDNN (base) model is an off-the-shelf system from the SpeechBrain toolkit \cite{ravanelli2021speechbrain}, trained on the VoxCeleb dataset \cite{nagrani2020voxceleb}, which primarily contains adult speech. The ECAPA-TDNN encoder produces 192-dimensional speaker embeddings from 80-dimensional log Mel filterbank features extracted from the input audio. We trained the ECAPA-TDNN model on the EDSI dataset using the same two-stage training procedure as the WavLM-TDNN model, and compared their performances under this training setup.

  \begin{table}[htbp]
\caption{Comparison of SV models across five folds using EER (\%). Avg: average EER across five folds. The best results are shown in bold.}
\footnotesize
\setlength{\tabcolsep}{2.2pt}
\renewcommand{\arraystretch}{1.4}
\centering
\begin{tabular}{|l|ccccc|c|}
\hline
 Model &Fold 1 %(Hawkins & Zazycki &Snipes &Cacaci)  
 &Fold 2 %Leggett&Cacaci & Kerr&Hawkins 
 &Fold 3 %Loux&Leggett&Paul&Snipes
 &Fold 4 %Snipes&Zazycki&Legget  &Britt
 &Fold 5 %Kerr&Hawkins&Crigger  &Paul
 &Avg
 \\
\hline 
 ECAPA-TDNN (Base)  &19.48&17.51 &18.80&21.75&23.75&20.26\\
\hline
ECAPA-TDNN &15.08&12.94 &16.81&17.86&19.49&16.44\\
\hline
WavLM-TDNN&\textbf{14.14}&\textbf{11.84}&\textbf{16.01}&\textbf{16.62}&\textbf{18.41}&\textbf{15.40}\\
\hline
    \end{tabular}
    %\vspace{-1mm} 
  \end{table}
  
As shown in Table 2, the WavLM-TDNN model consistently outperforms both the ECAPA-TDNN (base) model and the ECAPA-TDNN model trained on the EDSI classrooms dataset across all five folds, achieving average relative EER reductions of 23.99\% and 6.32\%, respectively. These results suggest that the proposed WavLM-TDNN architecture outperforms the ECAPA-TDNN model for SV in challenging classroom environments, where the diverse speech characteristics of children and adults, together with babble noise, increase task difficulty. 

Comparing the ECAPA-TDNN model fine-tuned on the EDSI classrooms dataset with the ECAPA-TDNN (base) model shows consistent EER reductions across all folds, achieving an average relative reduction of 18.85\%. This highlights the importance of domain adaptation for the SV task in classrooms. Training on classrooms data, which includes both child and adult speech as well as the acoustic challenges of real-world classroom interactions, such as babble noise from group discussions, leads to improved performance compared to a model trained primarily on adult speech.

%'Cunningham', 'Beverly', 'McMillan', 'Bagwoh', 'Britt', 'Snipes','Kerr', 'Zazycki', 'Cacaci', 'Cunningham' ,'Hawkins','McLeod','Borden','Paul','Shumaker','Williams','Leggett','Loux','Morgan','Crigger'
% fold1:'Zazycki','Snipes', 'Cacaci','Hawkins'
% fold2:"Leggett", "Britt","kerr" , 
% fold3: "Loux", "Leggett", "Paul", "Snipes"
% fold4: "Loux", "Zazycki", "McLeod","Britt"
% fold5:
\subsection{Evaluation of the WavLM-TDNN on multi-lingual classrooms data}
To extend the evaluation of SV models in classroom environments beyond English-only data, we conducted an experiment using a multi-lingual classroom session with speaker identity annotations, in which students spoke both Spanish and English. This dataset comprises 120 utterances with a total duration of 29 minutes. In this experiment, the WavLM-TDNN and ECAPA-TDNN models were trained on the 18 English-speaking classrooms dataset following the two-stage training approach. The models were then evaluated on the multi-lingual classroom data to assess their ability to generalize across languages. Since the models were trained exclusively on English speech, this setup provides an evaluation of cross-lingual generalization and robustness, demonstrating how well the SV models can handle classroom scenarios that include multiple languages. 

As shown in Table 3, the WavLM-TDNN model consistently outperforms both the ECAPA-TDNN (base) model and the ECAPA-TDNN model trained on the EDSI dataset, achieving relative EER reductions of 29.16\% and 5.53\%, respectively. These results demonstrate that WavLM-TDNN not only maintains strong performance on English speech but also generalizes effectively to non-English classroom speech, capturing speaker characteristics across languages. The cross-lingual performance of WavLM-TDNN highlights the benefit of combining self-supervised pre-trained speech representations with TDNN-based temporal modeling for SV tasks.

\begin{table}[htbp]
\caption{Comparison of SV models on multi-lingual classrooms data using EER (\%). The best result is shown in bold.}
\footnotesize
\setlength{\tabcolsep}{2.2pt}
\renewcommand{\arraystretch}{1.4}
\centering
\begin{tabular}{|l|c|}
\hline
 Model & Multi-lingual classroom
\\
\hline 
 ECAPA-TDNN (Base)  & 22.43\\
\hline
ECAPA-TDNN  &16.82\\
\hline
WavLM-TDNN&\textbf{15.89}\\\hline

    \end{tabular}
    %\vspace{-4mm} 
  \end{table}
\section{Conclusions and future work}
In this study, we investigate speaker verification (SV) in real English-speaking classroom settings, which include both children and adults and challenging acoustic conditions such as babble noise during group discussions. We demonstrate the effectiveness of a two-stage training strategy for SV, in which the model is first pre-trained using the momentum contrast (MoCo) self-supervised learning (SSL) framework and subsequently fine-tuned with a limited amount of labeled speaker data. This approach yields superior performance in classroom SV tasks compared to training with MoCo-based SSL alone. Furthermore, our results shows that the proposed WavLM-TDNN model trained under the two-stage framework outperforms the ECAPA-TDNN architecture on SV tasks. Evaluation on multi-lingual classroom data further shows that WavLM-TDNN consistently surpasses ECAPA-TDNN, demonstrating generalization beyond mono-lingual English-speaking classroom environments. 

Future work will focus on collecting additional classroom recordings with varied enrollment styles, such as read and spontaneous speech, to evaluate their impact on SV performance. Additionally, we plan to explore various SSL methods within the two-stage training framework and compare their effectiveness for SV in classroom settings.
%\newpage
\bibliographystyle{IEEEtran}
\bibliography{mybib}

\end{document}